To appear in *Proceedings of IEEE Symposium on Computational Intelligence for Financial Engineering (CIFEr),* Bengaluru, India; November 18-21, 2018.

# Deep Learning can Replicate Adaptive Traders in a Limit-Order-Book Financial Market

Arthur le Calvez[+] and Dave Cliff[*]
*Department of Computer Science*
*University of Bristol*
Bristol BS8 1UB, UK
[+]al14835@my.bristol.ac.uk, [*]csdtc@bristol.ac.uk

*Abstract*— We report successful results from using deep learning neural networks (DLNNs) to learn, purely by observation, the behavior of profitable traders in an electronic market closely modelled on the limit-order-book (LOB) market mechanisms that are commonly found in the real-world global financial markets for equities (stocks & shares), currencies, bonds, commodities, and derivatives. Successful real human traders, and advanced automated algorithmic trading systems, learn from experience and adapt over time as market conditions change; our DLNN learns to copy this adaptive trading behavior. A novel aspect of our work is that we do not involve the conventional approach of attempting to predict time-series of prices of tradeable securities. Instead, we collect large volumes of training data by observing only the quotes issued by a successful sales-trader in the market, details of the orders that trader is executing, and the data available on the LOB (as would usually be provided by a centralized exchange) over the period that the trader is active. In this paper we demonstrate that suitably configured DLNNs can learn to replicate the trading behavior of a successful adaptive automated trader, an algorithmic system previously demonstrated to outperform human traders. We also demonstrate that DLNNs can learn to perform better (i.e., more profitably) than the trader that provided the training data. We believe that this is the first ever demonstration that DLNNs can successfully replicate a human-like, or super-human, adaptive trader operating in a realistic emulation of a real-world financial market. Our results can be considered as proof-of-concept that a DLNN could, in principle, observe the actions of a human trader in a real financial market and over time learn to trade equally as well as that human trader, and possibly better.

## I. INTRODUCTION

The work described in this paper is a first step toward answering the following question: *is it possible to use contemporary machine-learning techniques, such as deep learning neural networks (DLNNs), to automatically learn to replicate the trading behaviour of a human trader in a financial market, purely by observing the actions of that trader in the market?* Human traders are valued for their ability to adapt, to learn from their experiences in the market, and so at the core of our question is whether DLNNs can learn to emulate a trader as he or she (or it) adapts to changing market conditions. We report here our preliminary results which indicate that the core question can be answered positively; and we interpret that as an indication that, by suitably extending the methods used here, it may in time be possible to sit a DLNN "black box" alongside a skilled human trader working in an investment bank or fund-management company; for the DLNN box to learn to copy the trader's actions purely by observing the actions of the trader; and then for the human trader to be replaced by the black box. If one human trader can be replaced then in principle all can be, and this thereby offers the potential for a considerable reduction in trading personnel.

In the past decade there has been a steep rise in the deployment of automated trading systems in major financial markets such as those for equities, currencies, bonds, commodities, and derivatives. Traditionally, for well over a hundred years, the work now done by automated trading system was instead the responsibility of human traders working initially face-to-face on the open-outcry trading floors of centralised exchanges; then via telephone communication; and more recently via messaging over computer networks. Human traders were typically very well paid for their work, which might charitably be interpreted as a sign that they were being appropriately rewarded for the intelligence required of them to be successful traders, rather than that they were simply lucky to have found themselves a job where pay levels were irrationally generous in comparison to the degree of intelligence required to successfully do that job. Thus, replacing human traders with automated systems is an interesting problem for the application of computational intelligence in financial engineering (CIFEr). A specific human job, known as a *sales trader*, involved working orders on behalf of a client: the sales trader does not hold inventory on her own account, but instead she attempts to execute buy or sell orders as specified by clients. A good sales trader tries to get the best price for the client, in return for a commission or fee. For example, if a client-order instructs the sales trader to buy Apple stock (ticker symbol AAPL) at a price of no more than $160, then if the sales trader can complete the order when AAPL is at $155, that's better than buying when AAPL is at $159. Similarly, if a client instructs the trader to sell IBM stock for no less than $150, it's better to execute that trade at a price of $160 than at $155. In both buying and selling, the difference between the price specified by the client (which we refer to here as the *limit price*) and the price at which the transaction is executed (the *transaction price*) can be thought of as "profit", either retained entirely by the sales trader, or shared in some ratio with the client. The trading floors of investment banks used to be the home of well-paid sales traders busily working away, either executing client orders or waiting for the next order to come in. And, in the past 15 years, almost all human sales traders working in spot markets (i.e., working orders for transactions that execute immediately) have been replaced by machines, by automated *algorithmic trading* systems, referred to in the industry as *robot traders* or simply as *algos*.

A key paper in the rise of algorithmic trading was written by a group of researchers at IBM's TJ Watson Research Lab, and presented at the 2001 *International Joint Conference on Artificial Intelligence* (IJCAI) [13]. This paper presented the first ever demonstration that automated trading systems could consistently outperform human traders. The IBM team ran a series of controlled laboratory experiments in which human traders interacted with one another in a simulated market that captured key aspects of real-world financial markets, with the interactions being mediated via a network of computer terminals – all interactions could be recorded with fine-resolution timestamps, which greatly eased subsequent analysis of the dynamics of the market experiments. This style of experiment was one that had been pioneered by the economist Vernon Smith (see e.g. [30]), who was later awarded the 2002 Nobel Prize in Economics for his leadership in establishing this approach, now widely practiced, and known as *Experimental Economics* (see also e.g.: [21, 22]). The key innovation in the IBM IJCAI paper was this: whereas Smith and other experimental economists had for many years been using networks of computer terminals to study the market-trading behaviour of human subjects; and while there was also a growing body of work in agent-based computational economics (see e.g.: [28] [34]) that explored the dynamics of markets populated entirely by robot traders; the IBM paper was the first to study under rigorous laboratory conditions the dynamics of interactions between human traders and robot traders. Specifically, the IBM team explored the interactions between human traders and two *adaptive* algorithmic trading systems (i.e., systems that could learn from their experience trading in the markets): one was a modified version of an adaptive probabilistic trading strategy originated by Gjerstad & Dickhaut [19], which IBM christened the *Modified Gjerstad-Dickhaut* (MGD) strategy; the other was the *Zero Intelligence Plus* (ZIP) adaptive trading strategy, which uses the same underlying machine learning mechanism as back-propagation neural networks (and hence also as that in many DLNNs); the complete source-code for ZIP had been published by Hewlett-Packard Laboratories four years earlier [7], and results from automated optimization of ZIP traders were presented at the 1998 IEEE CIFEr conference [8]. IBM's results showed that both MGD and ZIP could consistently out-perform human traders, in what was essentially a sales-trader role: i.e., MGD and ZIP robots could trade more profitably than human sales traders in financial markets. IBM's result generated world-wide media coverage.

It is beyond the scope of this paper to offer a summary of the history of events from publication of the IBM IJCAI paper to the current near-total adoption of automated execution systems in all of the world's major financial markets, and the consequent severe reduction in the number of human traders at the point of execution in those markets. For an historical overview of the rise of automated trading in financial markets, published by the UK Government's Office for Science, see [10]; and for commentaries on the negative aspects of markets with high degrees of adaptive automation see [1, 2, 9, 25, 27].

In the 17 years since the IBM paper was published, a number of other authors have replicated and extended their results: see e.g. [4, 5, 14-16]. To be realistic approximations of real-world financial markets, such experimental studies almost always employ a model of a limit-order book (LOB), a data structure very commonly shown on real trader's screens and commonly referred to as simply *the book*. Described in more detail below in Section II.B, the LOB presents key real-time information that typically all sales traders in a market rely on.

While algorithmic trading systems have replaced many human traders, especially sales traders in spot markets, there are still many humans employed to work in more complex and demanding trading roles. And that brings us back to the question posed at the start of this paper: the work described here is a first step at assessing whether, instead of hand-designing adaptive algorithmic trading strategies such as MGD and ZIP that outperform human traders, is it instead possible to use contemporary machine-learning techniques such as DLNNs to automatically learn to replicate the trading behaviour of an adaptive trader, purely by observing the actions of that trader in the market. Both human traders and algos such as MGD or ZIP are adaptive; in this paper we concentrate solely on exploring the ability of DLNNs to learn to capture the trading behaviour of adaptive algos. The alternative, running experiments with human subjects, incurs considerable costs in time and money, and very many experiments would be required to generate enough data to satisfy the requirements of DLNNs. So, as an initial proof of concept, we explore in this paper only the ability of DLNNs to learn to replicate the behaviour of an adaptive algorithmic trading system: specifically ZIP, which is known to outperform humans. We demonstrate here that DLNNs can indeed learn to replicate the live-trading behaviour of ZIP, and to our surprise we also learnt that in fact a DLNN trained only on data from ZIP behaviour can learn to perform consistently *better* than ZIP. Because ZIP has independently and repeatedly been demonstrated to out-perform human traders, if a DLNN system can outperform ZIP traders that that gives rise to the possibility that a DLNN system observing a human trader in a real financial market may be capable not only of learning to match that human trader's abilities, but also to exceed them.

The rest of this paper is structured as follows. In Section II we describe our methods and the key components of our proof of concept: our experiment platform is a system called BSE, an open-sourced minimal simulation of a LOB-based financial market; we populate BSE markets with a number of algo traders and run a large number of market experiments to generate trading data; that training data is then fed into the public-domain DLNN *TensorFlow* DLNN software library [33], via the *Keras* high-level toolkit [23] that provides abstractions and interfaces to ease use of TensorFlow for common application styles. Section III then presents results from our experiments, which demonstrate that DLNNs can indeed learn, purely by observation, to replicate the adaptive trading behaviour of a ZIP algo, and that the DLNN trader can be more profitable than the ZIP trader from which the DLNN's training data was generated: i.e., the DLNN "student" comes to outperform its ZIP "teacher". Section IV discusses further work prompted by the results described here, and our conclusions are offered in Section V.

## II. METHODS, COMPONENT TOOLS, AND RELATED WORK

For readers unfamiliar with DLNNs and the LOB real-time data-structure, brief introductions are given in Sections II.A and II.B respectively. Next, in Section II.C we present an outline of our method for training DLNNs to copy a sales trader's observable activity, a method that applies regardless of whether the trader being copied is a human or a robot/algo. As the time and money costs of attempting to do this with a real human trader in a real financial market would be nontrivial, and because this is an initial proof-of-concept study, we report here on the successful use a DLNN learning to copy the behaviour of an adaptive sales trader robot, specifically a ZIP trader, operating in a simulated LOB-based market that closely models key aspects of the LOB-based real-world financial markets. The LOB-market simulator is called BSE, described in Section II.D, and its use in the context of our experiments is explained in Section II.E. Finally Section II.F discusses the small number of publications by other authors that are relevant to the work reported here.

### A. Background: Deep Learning Neural Networks (DLNNs)

It is beyond the scope of this paper to explain in detail the operation of deep learning neural networks (DLNNs), a relatively novel style of machine learning algorithm that has proven to be highly successful in recent years. For the purposes of this article the reader does not require a detailed knowledge of how DLNNs work, other than the elementary information that they are based on parallel distributed processing architectures, each DLNN being a network of interconnected nodes separated into a number of layers, with a specific number of nodes in each layer, and with weighted connections between nodes. Each node in a DLNN implements a simple computation, typically producing an output value that is the result of a nonlinear function computed on the sum of weighted input values to that node. In DLNNs the values of input variables are first fed, modulated by the weight on each connection, into an initial *input layer* of nodes; the outputs of the input-layer nodes are fed as inputs into the next layer of nodes, referred to as the first *hidden layer*. The outputs from a layer of hidden nodes are then fed as inputs into the nodes in the next layer, which may be another hidden layer, or it may be a final layer of one or more *output nodes*, known as the *output layer*. Thus a DLNN may be so shallow as to only have three layers (the input layer, one hidden layer, and then the output layer), or it may be deeper if it has more than one hidden layer. In the DLNNs considered here, all of the weighted connections run from nodes in one layer to the next layer down in the direction of the output layer. That is, there are no feedback connections from nodes in one layer to nodes in earlier layers, and there are also no connections between the nodes in any one layer, i.e. we are working only with "feedforward" DLNNs.

DLNNs are most often used for so-called *supervised learning*, where a set of *training data* is repeatedly presented to the DLNN, and a learning algorithm adjusts the weights on the inter-node connections after each presentation of an item of data from the training set. Data items in the training set will typically consist of a pair of vectors: one vector of input values and one "target" vector of desired output values that the network should produce when the associated input vector of values is fed to the DLNN's input nodes. Over time, the learning algorithm acts to reduce the error between the target output and the actual output of the network for each input vector in the training set. After the errors have been reduced or minimised by the learning algorithm, the extent to which the DLNN has usefully learned to generalise the mapping from input vectors to output vectors is then evaluated by exposing it to a set of *test data*, typically drawn from the same distribution as the training data. If the DLNN has learned an appropriate mapping, the error on the test-set should not be significantly higher than the error achieved on the training set at the end of the learning process. Typically, whether a DLNN learns a useful representation of the mapping from input vectors to output vectors is dependent on a number of factors, hyper-parameters of the learning system. Particularly important hyperparameters are those that determine the network architecture, i.e. the number of layers and the number of units in each layer: setting appropriate values for these numbers is often more of an art than a science, involving a process of trial and error, and educated guesses: a topic we return to in Section IV. For further details of DLNNs, see e.g. [24, 20, 6, 18, 26].

### B. Background: The Limit Order Book (LOB).

In almost all of the world's major financial markets, buyers and sellers interact via a mechanism known technically as the *Continuous Double Auction* (CDA). In the CDA, any buyer can announce or *quote* a bid at any time, and any seller can quote an offer at any time. While this is happening, any seller can accept any buyer's bid at any time; and any buyer can accept any seller's offer at any time. It's a continuous asynchronous process, and it needs no centralized auctioneer, but it does need some way of recording the bids and offers that have been made and not yet transacted: this record is the *limit order book* (LOB). In financial-market terminology, a *limit order* is one that will only be executed when a counterparty is found who is happily to transact at that order's pre-specified *limit price*: this distinguishes it from other types of order that execute immediately, for example an *at-market order* which clears by taking whatever price the market will bear at the moment the order is submitted.

The LOB is a data-structure published, with real-time updating, as the primary summary of the current state of the market supply and demand for a particular tradeable item (e.g. a specific stock, a specific currency pair, a specific fixed-income bond, etc). Almost every electronic financial market exchange publishes a continuously-updated LOB for each security traded on that exchange. In markets such as foreign exchange (FX) there are no formally instituted centralised exchanges, but aggregation services publish real-time FX price streams that can readily be displayed in a GUI as a close analog of a LOB.

The CDA has been the subject of much study in economics and finance (see, e.g. [17]). It interests economists because, even with a very small number of traders, the *transaction prices* (i.e. the agreed deal-prices) rapidly approach the theoretical market equilibrium price. The equilibrium price is the price that best matches the quantity demanded to the quantity supplied by the market, and in that sense it is the most efficient price for the market. The CDA is also of pragmatic interest because of the trillions of dollars that flow through national and international CDA-based markets around the world each day. Although there are still some exchanges where human traders physically meet in a central trading pit and shout out verbal bids and offers, in very many major markets the traders engage with one another remotely, via a screen-based electronic market, interacting by placing quotes for specific quantities at specific prices on the LOB.

The LOB displays data that summarizes all the outstanding bids and offers, i.e. the "live" orders that have not yet cancelled by the traders that originated them. In market terminology, offers are also known as *asks*, and the LOB has two sides: the bid side and the ask side. The bid side shows the prices of outstanding bid limit orders, and the quantity available at each of those prices, in descending order of price, so that the best (highest) bid is at the top of the book. The ask side shows the prices of outstanding asks, and the associated quantities, in ascending price order, so that the best (lowest) ask is at the top.

In financial-market terminology, the *spread* is the difference between the best ask and best bid. If a trader wants to sell at the current best bid price, that's referred to as *hitting the bid*; if a trader wants to buy at the current best ask-price, that's referred to as *lifting the ask*. Both hits and lifts can be signaled by the trader issuing a quote that *crosses* the spread, i.e. issuing a bid priced at more than the current best ask, or issuing an ask priced at less than the current best bid: the transaction then goes through at whatever the best price was on the LOB as the crossing quote was issued; the price on the crossing quote is irrelevant, so long as it crosses the spread.

So, for example, if there are two traders each seeking to buy 30 shares in company XYZ for no more than $1.50 per share, and one trader hoping to buy 10 for a price of $1.52; and at the same time if there was one trader offering 20 shares at $1.55 and another trader offering 50 shares at $1.62, the LOB for XYZ would appear as illustrated in Fig.1, and traders would speak of XYZ being priced at "152-55" with the spread being $0.03, and the *midprice* being the arithmetic mean of the best bid price and best offer price.

The information shown on a LOB is sometimes referred to as "Level 2" or "market depth" data. In contrast, "Level 1" market data just shows the price and size (quantity) for the best bid and ask, along with the price and size of the last recorded transaction for the instrument being traded. Some people like to try their hand at "day trading" on their home PCs and they often operate with even more restricted data, such as the time-series of whatever price the instrument was last traded at, or the mid-price, the point between the current best bid and the best ask. Typically the richer the and more voluminous the financial market data, the more expensive it is to purchase from a commercial provider. Full Level 2 data is routinely used by professional traders in investment banks and hedge funds, but researchers in those institutions are famously much better resourced than meagerly-funded university academics. Buying licenses for historical records of Level 2 data in the large quantities needed for DLNN training is unfeasibly expensive at this early exploratory proof-of-concept stage of our research: a cheaper alternative to buying Level 2 historical data is needed, which is why we used instead the LOB-market simulator, BSE, described in Section II.D below.

| XYZ | | | |
|---|---|---|---|
| **Bid** | | **Ask** | |
| 10 | 152 | 155 | 20 |
| 60 | 150 | 162 | 50 |

Fig.1: How a Limit Order Book (LOB) for a fictional tradeable security with ticker-symbol XYZ might appear on a trader's screen. The bid-side of the book appears on the left, with highest prices at the top; the ask-side appears on the right, with lowest prices on the top. The best bid is $1.52, with 10 units of XYZ demanded at that price; the best ask price is $1.55, with 20 units offered at that price. Both sides of the book have a *depth* of 70, the bid-ask *spread* is £0.03, and the market *midprice* is $1.535. See text for further explanation

*C. Method: Training a DLNN by Observing a Sales Trader*

If we were working on attempting to have a DLNN learn to replicate the adaptive trading behavior of a real human sales-trader in a real-world LOB-based financial market, on the assumption that the trader is using a trading terminal rather than trading via voice phone-calls, it would be trivially easy to record the orders that the trader is being asked to execute, as they come in from the clients, and to also record the limit-order quotes (bids or asks) that the trader is issuing to the market via her trading terminal. Even if she is making voice calls, speech recognition software could be used to capture the stream of quotes and responses in the telephone call. Assuming that an accurate timestamp is associated with each quote issued by the trader; and that a timestamp is also associated with each order as it comes in from a client; and furthermore assuming that each time the LOB changes, that change is also logged on a timestamped data record, then it is straightforward to build a "consolidated tape", i.e. a single time-series of events accurately recording the sequence of events involving arrival of orders to the trader, the issuing of quotes by that trader, and the changes in the LOB over some trading period. This tape of data could be divided into a set of training data and a set of test data. The test data would be used while the DLNN is learning, adjusting its weights to reduce errors; and the test set would then be used to evaluate the usefulness of the trained DLNN after the learning process terminates.

Although the work reported here does not involve taking data from a real human trader in a real-world financial market, because of the potentially prohibitive costs of doings so as discussed earlier in this paper, exactly the same method as sketched above can be applied in our proof of concept, attempting to have a DLNN learn to replicate the adaptive trading behaviour of a ZIP sales-trader in a suitably realistic simulation of a LOB-based financial market. To most clearly explain how we did this, Section II.D first introduces *BSE*, the LOB-market simulation that we used as the platform for our experiments, and then Section II.E provides further details on how we used BSE, along with the public-domain machine-learning DLNN software toolkits *TensorFlow* and *Keras*, to set up our machine learning experiments; and Section II.F then closes with a discussion of related work.

*D. Apparatus: the BSE minimal simulation of a LOB-market*

What is needed, ideally, is a real-world financial exchange to treat as an experiment test-bed, but that is clearly not a feasible prospect. However a viable alternative is to use a simulation of a LOB-based market, that adequately captures the relevant details for the purposes of this research: BSE (the Bristol Stock Exchange) is an open-source simulator that meets those needs. BSE is described in some detail in [11, 12]. In summary, its key features are:

- While real financial exchanges will typically simultaneously maintain LOBs for tens, hundreds, or thousands of types of tradeable item (i.e., different stocks, different currency-pairs, or different commodities), BSE has just one LOB, for recording limit orders in a single anonymous/abstract type of tradeable item.
- BSE allows for the experimenter to control the specification of any of a wide range of dynamics of supply and demand in the BSE market.
- BSE includes a number of pre-coded robot trading algorithms drawn from the literature on automated trading over the past 30 years, including ZIP. This allows the user to explore the dynamics of LOB-based CDA markets without having to write their own robot-trading algos.
- BSE's pre-coded robot trading algorithms can easily be copied and amended/edited/replaced, giving users of BSE the opportunity to explore developing their own robot trader algorithms and to evaluate them against various combinations of the stock of existing algos in BSE.

BSE is deliberately written as a simple, intelligible, single-threaded minimal simulation, in the widely used programming language Python (Version 2.6). The code has been available as open-source on the GitHub repository since October 2012 and has been downloaded many times, with a number of individuals having spawned forks and uploaded extensions of the system, including the addition of Vytelingum's *Adaptive Aggressive* (AA) robot trader strategy [37], which was demonstrated in a 2011 IJCAI paper [14] to be the most dominant adaptive robot-trader strategy when pitted against human traders. For further details of BSE, see [11, 12].

It is important to remember that BSE as configured here is a *minimal* simulation of a financial exchange running a LOB in only a single tradable security. It abstracts away or simply ignores very many complexities that can be found in a real financial exchange. In particular, we used BSE configured in such a way that a trader can at any time issue a new order, which replaces any previous order that the trader had on the LOB: that is, any one trader can have at most one order on a LOB at any one time; also we used BSE set up so that all orders were for a quantity of one. This is consistent with long-established norms and practices in experimental agent-based economics. Furthermore, as currently configured, BSE assumes zero latency in communications between the traders and the exchange, and also (very conveniently) assumes that after any one trader issues an order that alters the LOB, the updated LOB is distributed to all traders (and, potentially, also results in a transaction) before any other trader can issue another order. Real financial exchanges are significantly more complex than this, particularly with regard to latencies.

*E. Using BSE as a Source of Sales-Trader Training Data*

The method described in broad terms in Section II.C was implemented as follows. BSE was used to create markets populated entirely by robot traders, and various schedules of supply and demand were set up to stochastically generate, from nonstationary (i.e. time-dependent) distributions, client-orders to sell or buy units of the one arbitrary tradeable security in BSE. The schedules could be set up to give "bull" markets with steadily rising prices, "bear" markets with steadily declining prices, stagnant markets where prices take a drunkards walk along a flat line, and markets with various forms of oscillation or multi-phasic variation between periods of bull runs, bear runs, and sideways movement. A fresh randomly-generated client-order was fed to each trader as it completed its last assignment from a client. The open-source version of BSE [11] was edited (in straightforward ways) so that a "consolidated tape" (as introduced above in Section II.C) recording timestamped LOB data, client-order limit-price, and quote activity was produced for each trader in the market. After a number of market sessions, each of which involved several thousand orders, a successful ZIP trader would be identified and data from its consolidated tape was then divided into a training set and a test set, which were then used for training and evaluation of the DLNN trader.

As explained fully in [3], the responses of two types of DLNN were explored: Multi-Layer Perceptron [24, 6, 26]; and Long-Short-Term Memory [20, 6, 26]. We refer to these two types as MLP-DLNN and LSTM-DLNN respectively. These had previously been explored in earlier work at Bristol [35], discussed in Section II.F below, and network architectures that had shown promise in [35] formed the starting point for the work reported here. Root mean square (RMS) error was recorded for the training set and the test set, and those DLNNs for which both errors were sufficiently low after training were considered successful. Both the MLP and the LSTM DLNN were created via the open-source toolkit *Keras* [18] running on top of the open source machine learning software system *Tensorflow* [33, 23]. Sample configuration files for Keras, used in generating the results reported here, are presented in [3].

Successful DLNNs were then added into the BSE market, alongside the other robot traders, and participated in the market activities for sufficiently long to generate reliable results from "live trading", the ultimate test of whether the DLNN trader has truly learned to replicate (or improve upon) the performance of the original adaptive trading algorithm that generated the training and test data. Two forms of live-trading were explored: *heterogenous* where a single trained DLNN trader was added back into the market from which that trader's training data had been generated; and *homogeneous* where a market was populated entirely by clones of the trained DLNN trader: such homogenous markets have routinely been studied in earlier work (e.g. [7, 8, 37]).

## F. Related Work

Although there is a long-established and very large literature on time-series forecasting in general, and the prediction of financial time series in particular, it is important to appreciate that none of that work is relevant here because we are not using machine learning to predict future transaction prices in a market. Instead, we are using machine learning to replicate the sequence of actions generated by a trader in response to the dynamically changing LOB; and what actions are appropriate are context dependent, being determined by that trader's current limit-price (as specified in the client-order that the trader is currently working). Whether the quote issued by the DLNN trader is accepted by another trader or ignored will determine whether that particular quote has any effect on the time series of transaction prices in the market or not, but trades that are ignored (and hence have no direct effect on the transaction-price time series) can nevertheless play a significant role in price-discovery, i.e. in altering the array of bids and offers on the LOB until a trade does occur.

To the best of our knowledge, there are only three directly relevant pieces of prior work, all of which are so recent that they are currently unpublished or in press, and none of which present results directly comparable to ours.

The first is a 2017 Master's Thesis [35], supervised by one of us (Cliff), which made some promising preliminary progress in using DLNNs to replicate the behaviour of ZIP traders in BSE, but did not make the crucial step of subsequently "live trading" the trained DLNN, as described previously in Section II.E.

The second is a journal paper [36], accepted for publication in the *IEEE Transactions on Industrial Informatics* but currently still in press, presumably to appear later in 2018. In this paper LSTM is used to replicate the elementary behaviour, in the form of generating simple buy/hold/sell trading signals, resulting from three very simple non-adaptive trading strategies that work purely on a simple transaction-price time-series and hence do not use Level 2 LOB data. The three strategies are: *Random Choice* (which pays no attention to the financial time series); *Crossover* (which generates trading signals if/when a short-term moving average of the time-series crosses over a longer-term moving average); and *MACD* (which stands for moving-average convergence/divergence, an extension of the Crossover strategy that includes an oscillating momentum term). The authors of [36] report success with their methods, which is evidence that DLNN can learn simple non-adaptive trading strategies. However our work significantly extends their result, because we use LOB data and require the DLNN to learn the trading behaviour of an adaptive robot, one that learns from its experience in the market rather than blindly following the same simple strategy regardless of market events.

The third publication [29] is currently available only as a preprint on the popular *ArXiv* website, and is presumably currently undergoing peer-review. In this paper, the authors report on using DLNNs trained on truly vast quantities of real-world high-frequency financial-market quotes and transactions. The results in [29] demonstrate that using price and order-flow history can improve forecasting performance and reveal evidence of path-dependencies in the price dynamics of real-world financial markets. Unlike our work, there is no attempt in [29] to explicitly learn to replicate the behavior of a specific adaptive trader (human or robot), but rather to gain insights into reliably observable statistical regularities within the billions of items of quote and transaction data.

And so, to the best of our knowledge, the results presented here are the first demonstration of using DLNNs to successfully replicate an adaptive trader, with capabilities that have previously been demonstrated to be comparable to a human trader, in a realistic (LOB-based) financial market.

## III. RESULTS AND ANALYSIS

Fig.2 shows illustrative successful results from a (MLP-DLNN trained on data from a ZIP trader working buy-orders for customers in a market where the transactions prices are trending down: the figure-caption explains the details of the graph. As is clear from the figure, there is a good qualitative match between the time-series of quotes from the ZIP trader, used in training MLP-DLNN, and the time series of quotes issued by the MLP-DLNN when it is presented with input vectors from the set of training data: this indicates that the learning has succeeded in capturing the mapping from input vectors to output vectors during training. The good qualitative match is maintained when the MLP-DLNN trader is presented with previously unseen training data, indicating that the DLNN system has learned a usefully nontrivial generalization, and has not over-fitted to the test-set. Fig.2 is representative of all the successful results generated: for a further 9 example graphs, in various types of market (bull, bear, sideways, multiphasic) see Fig 3.2 and 3.3 of [3], which also tabulates RMS error values.

Results from a heterogeneous live-trading test (as described above in Section II.E) of a successful MLP-DLNN are shown in Fig.3. This shows a good qualitative match between the ZIP "teacher" time-series of orders and the corresponding MLP-DLNN "student" time series, but what really counts in live trading is profit accumulated by the trader, i.e. *profitability*.

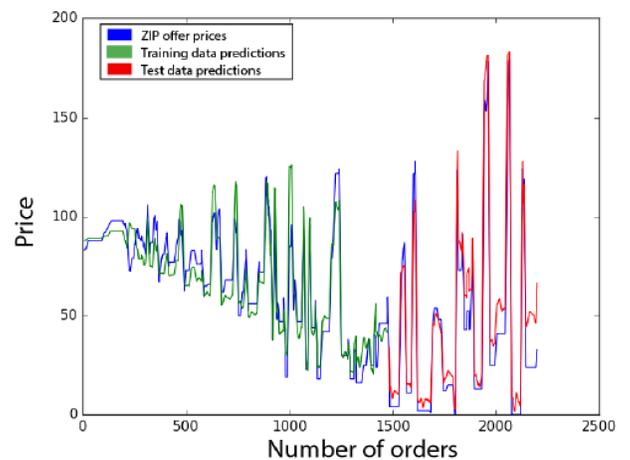

Fig.2: Illustrative results, the stream of quote-prices, from a MLP-DLNN trained on data from a ZIP robot trader working client buy-orders in a market where price is trending downwards, for a stream of c.2,200 quotes. The blue line is the time-series of offer prices generated by the ZIP trader operating live in BSE and recorded to produce a set of target output vectors used in training and testing. The input vectors (LOB data and the client-order limit price) are not shown. The green line shows the corresponding outputs from the DLNN working on input vectors used in the training set, and the red line shows the outputs from the DLNN responding to input vectors used in the test set.

To quantify the difference in profitability between ZIP traders and a trained MLP-DLNN trader, we first conducted 50 statistically independent runs of homogeneous market sessions where all 10 traders in the market were identical ZIP traders; and we then replaced one arbitrarily-chosen ZIP trader with the trained MLP-DLNN trader and ran another heterogenous 50 market sessions. In the heterogenous market sessions we recorded the final profit accumulated at the end of each session by the single MLP-DLNN trader, giving us 50 samples from that trader; and from the homogeneous ZIP markets we recorded the 50 end-of session profit values scored by the ZIP trader that was subsequently replaced by the MLP-DLNN trader. This gave us two samples of end-of-session profit values, each of size $n=50$. The box-and-whisker plot in Fig.4 shows comparison of key summary statistics from the two $n=50$ sample distributions, generated from the MLP-DLNN and ZIP traders that generated the time-series data in Figure 3.

A one-tailed (i.e., directional) Wilcoxon-Mann-Whitney U-test on the data illustrated in Fig.4 indicates that the MLP result is significantly better than the ZIP result: $p<0.01$; that is, the MLP-DLNN profits are statistically significantly better (i.e. higher) than the profits of the comparable ZIP-trader markets. Additional results presented in [3] demonstrate similarly significant (and larger-magnitude) performance advantages of MLP-DLNN traders over ZIP traders in heterogeneous markets, indicating that the results in Fig.4 are not an isolated success.

Because ZIP traders have previously been demonstrated to outperform human traders, and because MLP-DLNN traders have here been demonstrated to outperform ZIP traders, it seems reasonable to syllogistically infer that our results presented here demonstrate that MLP-DLNN traders can also learn to outperform human traders. That is, we have demonstrated here that deep learning can be applied to replicate the behaviour of adaptive traders in LOB-based auction markets, of the type used throughout the world's major financial markets, working from training data generated by observing the actions of algo traders already known to consistently outperform human traders.

For brevity, we have here relied upon a small number of illustrative graphs to make the case that MLP-DLNN "student" trader can learn to equal or outperform the performance of a ZIP "teacher" trader; additional graphs, tables, and statistical significance tests are presented in [3]. Nevertheless, as is manifestly clear from Fig. 4, MLP-DLNN learns to significantly outperform ZIP in the live-trading tests.

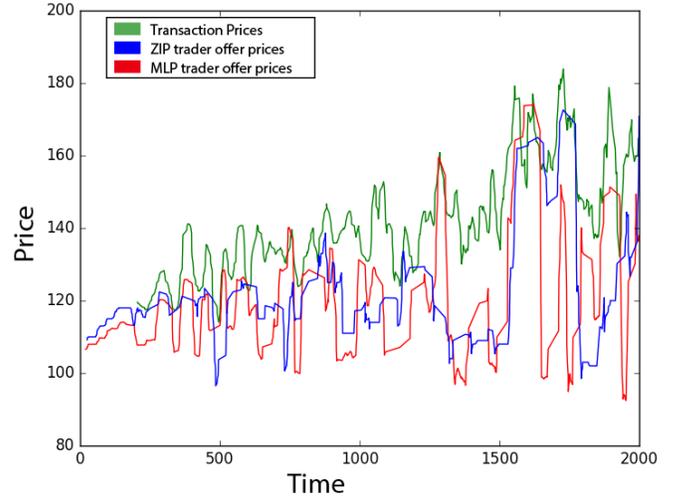

Fig.3: Illustrative results from a heterogeneous live-trading test of a successfully trained MLP-DLNN; i.e. where one well-trained MLP-DLNN is inserted back into the market from which its training and test data were generated. The time-series of transaction prices in the market is shown by the green line; the time-series of quote prices from the ZIP "teacher" automated trading system is shown in blue, and the time-series of quote prices from the "student" MLP-DLNN trader are shown in red. There is good qualitative agreement between the trading activity of the ZIP teacher and the MLP-DLNN student. Periods where transaction prices (green) are a long way from quote prices (red/blue) indicate this trader attempting to work an order in adverse market circumstances.

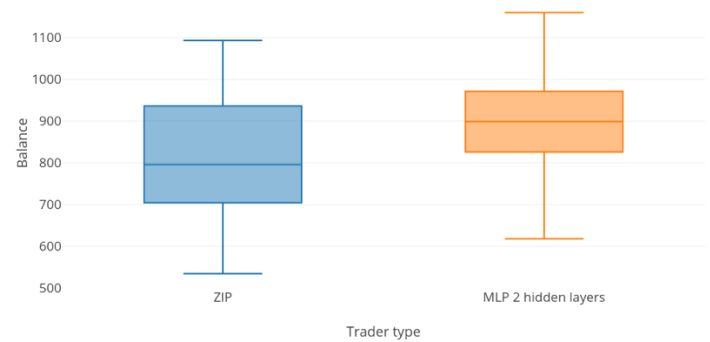

Fig.4: Summary statistics from 50 homogeneous market sessions with 10 ZIP traders (blue, left-hand box-and-whisker plot) and from 50 heterogenous market sessions with 9 Zip traders and 1 successful MLP-DLNN trader (orange, right-hand plot). The vertical scale is total accumulated profit at the end of the market session. The horizontal line within the shaded box is the median value in the sample of size $n=50$; the upper and lower bounds of the shaded box are the upper and lower quartiles, respectively. The whiskers end at plus and minus two standard deviations.

## IV. FURTHER WORK

As we take this work further, we are currently concentrating on two main avenues of enquiry. One is exploring the extent to which those DLNNs that successfully learn to outperform the trader used to generate the training data can be analysed, to reveal how and why they outperform the original trader, i.e. to come up with causal mechanistic characterisations and explanations for the improved performance. The other is to use automated optimization techniques to eliminate the trial-and-error human guesswork in arriving at a satisfactory network architecture (number of layers, number of nodes in each layer) for the DLNNs. Work is underway, with some promising early results, and we expect to report positive progress in future papers.

## V. CONCLUSIONS

In this paper we have demonstrated that a suitably configured DLNN can learn to replicate the trading behavior of a successful adaptive automated trader, the ZIP algorithmic strategy previously demonstrated to outperform human traders. We also demonstrate that DLNNs can learn to perform better (i.e., more profitably) than the ZIP trader that provided the training data. We believe that this is the first ever demonstration that DLNNs can successfully replicate a demonstrably better-than-human adaptive trader operating in a realistic emulation of a real-world financial market. Our results can also be considered as proof-of-concept that a DLNN could, in principle, observe the actions of a human trader in a real financial market and over time learn to trade equally as well as that human trader, and possibly better.